\documentclass[twocolumn,secnumarabic,amssymb, nobibnotes, aps, prd]{revtex4-2}

 \clearpage

\def\psl{\hbox{\hbox{${p}$}}\kern-1.9mm{\hbox{${/}$}}}
\def\dsl{\hbox{\hbox{${\partial}$}}\kern-1.7mm{\hbox{${/}$}}}
\def\Dsl{\hbox{\hbox{${D}$}}\kern-2.1mm{\hbox{${/}$}}}

\newcommand{\bp}{M_P}

\usepackage[table]{xcolor}

\usepackage{epstopdf}

\def\Lag{\mathscr{L}}

\def\be{\begin{equation}}
\def\ee{\end{equation}}
\def\bea{\begin{eqnarray}}
\def\eea{\end{eqnarray}}

\def\ba{\begin{array} }
\def\ea{\end{array}}
\def\bac{\begin{array} {c}}
\def\bacc{\begin{array} {cc}}
\def\baccc{\begin{array} {ccc}}

\usepackage{amssymb}
\usepackage{amssymb,exscale}
\usepackage{color}
\usepackage{xcolor}
\usepackage{mathrsfs}
\usepackage{epsfig}
\usepackage{mathrsfs}
\usepackage{amsfonts}
\usepackage{slashed}
\usepackage{blindtext}
 \usepackage{fancyhdr}
 \usepackage{mathtools}
 
\usepackage{comment}
\usepackage{graphicx}
\usepackage{amsmath}
\usepackage{mathtools}
\usepackage{slashed}
\usepackage{xcolor}
\usepackage[force]{feynmp-auto}
\usepackage{grffile}
\usepackage{float}
\usepackage{bbm}
\usepackage{hyperref}
\usepackage{ragged2e}
\usepackage{blindtext}

\allowdisplaybreaks

\begin{document}

\title{\LARGE ACT-Planck data and phase transitions \\ from a viable no-scale Standard Model completion}%

\author{{ Filippo Cutrona, Francesco Rescigno and Alberto Salvio} \\
\vspace{0.cm}
{\it Physics Department, University of Rome and INFN Tor Vergata, Italy}\\
\vspace{0.6cm}
\begin{abstract}
\noindent  Classically scale-invariant (and perturbative) theories provide a way to understand large hierarchies, as scales are generated through dimensional transmutation. They always lead to first-order phase transitions, since symmetries are radiatively broken, and they generically feature quasi-flat potentials, which are suitable for inflation. We construct a simple but fully realistic model of this kind that accounts for all observational evidence of new physics and is remarkably compatible with the most recent constraints on inflationary observables from both the Planck/BICEP/Keck and the Atacama Cosmology Telescope (ACT) collaborations. This model illustrates how classical scale invariance generically leads to a non-standard cosmology in which inflation occurs in two stages: a slow-roll stage and a thermal stage, separated by a radiation-dominated era.
\end{abstract}
}%



\maketitle


\tableofcontents

\section{Introduction}

Cosmology offers the possibility of testing particle-physics theories at energies much higher than those accessible at particle accelerators.  A classic example is inflation, which can involve energies a few orders of magnitude below the Planck scale. Moreover, after inflation, a period of radiation dominance takes place and the high temperatures reached during such an epoch can restore symmetries that are broken at zero temperature~\cite{Weinberg:1974hy}. When the temperature $T$ drops to a sufficiently low value, phase transitions (PTs) can  occur and those symmetries are broken again. Both inflation and PTs, if first order and sufficiently strong, can lead to observable consequences. They leave observable imprints in the cosmic microwave background (CMB) and can produce gravitational waves (GWs) that can be tested with current and/or future GW detectors~\cite{Maggiore:2018sht}. 

The Standard Model (SM) of particle physics is known to be incomplete due to well-established phenomena such as neutrino oscillations, dark matter, and the baryon asymmetry. A significant portion of these phenomena has a cosmological origin. It is therefore natural to use cosmological data to test SM extensions that address these observational issues.

Furthermore, the SM does not offer any explanation of the huge hierarchy between the Planck and the Fermi scales. SM extensions with no (fundamental) scales, a.k.a.~classically scale invariant (CSI) theories provide an explanation~\cite{Gildener:1976ih,Salvio:2025oem}. In these theories mass scales are generated through dimensional transmutation: the electroweak (EW) symmetry is broken radiatively (through quantum corrections) because mass scales are not present in the classical theory. In the general radiative-symmetry-breaking (RSB) scenario there must exist a flat direction of the effective potential at some RG scale value $\tilde\mu$  and a quartic coupling must then vanish at $\tilde\mu$~\cite{Coleman:1973jx,Gildener:1976ih,Salvio:2023qgb,Salvio:2026bco}. The flat direction is parameterized by a real scalar, which we call $\chi$. Dimensional transmutation can lead to exponentially large hierarchies. This is because the renormalization group (RG) running of a generic dimensionless coupling $\kappa$ is logarithmic in the RG (energy) scale $\mu$. So order-one variations of $\kappa$ correspond to exponentially large variations of $\mu$. This is why physicists do not consider the large hierarchy between the scale below which quarks are confined into hadrons and the Planck mass to be puzzling; rather, it is seen as a natural feature.

While dimensional transmutation has the potential to explain large hierarchies, it should be kept in mind that this mechanism does not automatically satisfies 't Hooft's  principle of naturalness~\cite{tHooft:1979rat} (the fact that a parameter is naturally small when taking it to zero leads to an enhanced symmetry). According to this definition, a naturally small parameter remains small when one takes into account quantum corrections coming from states heavier than the largest energy consistently described by the model (the cutoff). In order to satisfy such naturalness criterion, a model with dimensional transmutation should be embedded in a  theory where the smallness of the parameter under study (e.g.~the ratio between the Planck and the Fermi scales) is protected by a symmetry. 

\vspace{0.1cm}

{\it The goal of this paper is to identify and investigate a simple yet realistic CSI model. By ``realistic" we mean a model that is consistent with all experiments and observations, including those performed at colliders, the above-mentioned evidence of new physics and all available information on inflation.} 

\vspace{0.1cm}

We do not attempt here to embed this realistic CSI model into a more complete theory where 't Hooft naturalness is satisfied and large quantum corrections from the heavier states responsible for the UV completion of all interactions (gravity included) are avoided through extra symmetries.

It is also appropriate to mention that in our work scale invariance is a global symmetry. Constructions where scale symmetry is gauged, such as~\cite{Ghilencea:2018thl}, have redundant degrees of freedom that can be eliminated through a gauge fixing.

In Refs.~\cite{ACT:2025fju,ACT:2025tim} the  ACT collaboration provided the most recent determinations of several inflationary observables. One of the most evident differences with respect to the previous values given by the Planck and BICEP/Keck (BK18) collaborations~\cite{Ade:2015lrj,BICEP:2021xfz}  is a higher favored value for the spectral index $n_s$ of scalar perturbations: when the new ACT observations are combined with Planck and baryon acoustic oscillation data from the Dark Energy Spectroscopic Instrument (DESI)~\cite{DESI:2024uvr} (this combination is denoted here  Planck-ACT-LB) Ref.~\cite{ACT:2025fju} finds $n_s=0.9743 \pm 0.0034$.

 If one requires perturbativity, CSI theories predict a nearly-flat potential density for $\chi$, as long as a sizable CSI non-minimal coupling between $\chi$ and the Ricci scalar is present~\cite{Marzola:2016xgb,Kannike:2014mia,Marzola:2015xbh}. This feature can accommodate the most recent ACT values~\cite{Gialamas:2025kef,siACT,Kallosh:2025rni}. Furthermore, in the CSI scenario RSB is always associated with strong first-order PTs~\cite{Witten:1980ez,Salvio:2023qgb}. Evidence of first-order PTs in the form of GWs would provide further evidence of new physics (as the SM does not predict this type of transition in cosmological settings) and constitute an intriguing indication that the SM extension we are seeking is CSI.
 First-order PTs in the CSI scenario can also lead to primordial black holes (PBHs)~\cite{Liu:2021svg,Gouttenoire:2023naa,Salvio:2023ynn,Gouttenoire:2023pxh,Salvio:2023blb,Arteaga:2024vde,Ning:2026nfs}

\vspace{0.1cm}

{\it The paper is organized as follows. In Sec.~\ref{Building the model} one of the most economical, if not the most economical model, with all the above-mentioned properties is identified.  In Sec.~\ref{Inflation} we demonstrate that this model is compatible with the most recent inflationary observations. In Sec.~\ref{Reheating and thermal inflation} we show that, after an initial period of slow-roll inflation and the subsequent reheating and radiation dominance, another inflationary period occurs because of the first-order PT associated with EW symmetry breaking. The conclusions are in Sec.~\ref{Conclusions}.}

\section{Building the model}\label{Building the model}

One of the most economical ways to  account for neutrino oscillations, dark matter and baryon asymmetry is to add three right-handed neutrinos  featuring Majorana masses much below the EW scale~(see e.g.~\cite{Asaka:2005pn,Canetti:2012kh}). We thus introduce such neutrinos $N_i$ (with $i=1,2,3$).

Moreover, an extra scalar beyond the Higgs field is needed here because, as well-known, identifying the Higgs with $\chi$
would lead to a Higgs mass significantly smaller than the observed value. We thus introduce a scalar $A$. We take it with a non-vanishing lepton number to reduce the number of independent terms in the action and thus simplify the model.

The corresponding Majorana mass terms can then be promoted to scale-invariant Yukawa interactions
\be  \frac12 y_{ij} A N_iN_j + \mbox{h.c.}, \label{ANYuk}\ee 
where the $y_{ij}$ are the corresponding Yukawa couplings. 

The symmetries of the theory also allow for a portal interaction between the Higgs doublet $\mathcal{H}$ and $A$ of the form 
$\lambda_{ah} |A|^2|\mathcal{H}|^2$. This interaction and~(\ref{ANYuk}) are responsible, respectively, for the radiative breaking of the lepton symmetry and the EW symmetry when $A$ acquires a vacuum expectation value (VEV). This VEV has to be at a sufficiently high mass scale and  the portal coupling $\lambda_{ah}$   should be sufficiently small to fulfil the experimental bounds and  generate the observed value of the EW scale. A value of $\lambda_{ah}$ a few orders of magnitude below 1 is sufficient to this purpose. In this limit the flat direction should be, at least approximately, along $|A|$ and the quartic coupling $\lambda_\chi$ that vanishes should be, at least approximately, the quartic coupling $\lambda_a$ of $|A|$. 

With only $A$ and the $N_i$ as beyond-the-SM fields the effective coupling of $\chi$ to the other fields is too small for the bubbles not to be diluted by the expansion of the universe, at least if gravity is neglected in the nucleation process~\cite{Salvio:2023ynn}. Taking into account gravity, on the other hand, one could avoid this problem but, as we will discuss in Sec.~\ref{Reheating and thermal inflation}, a number of inflationary e-folds $N_e$ some tens larger than 60 is obtained (see e.g.~\cite{Liddle:2003as,WeinbergCosmo} for an upper limit on $N_e$).
 
 This inconvenient feature can be entirely avoided by gauging the Abelian U(1) symmetry acting on $A$. As well known, in order to avoid gauge anomalies, such new gauge symmetry must correspond to $B-L$. So we call it $U(1)_{B-L}$. Therefore, all leptons (including the $N_i$), all quarks and $A$ are charged under $U(1)_{B-L}$. Choosing the gauge coupling $g_1'$ of $U(1)_{B-L}$ sizable,   the bubbles are eventually not diluted by the expansion of the universe even  if gravity is neglected in the nucleation process~\cite{Salvio:2023ynn}. We will take such a value of $g_1'$ in this paper in a way that the effective coupling of $\chi$ is dominated by $g_1'$.

The matter Lagrangian is given by~\cite{indices}
\bea \Lag_{\rm matter}&=&\Lag^{\rm ns}_{\rm SM} + D_\mu A^\dagger D^\mu A + \bar N_j i\slashed{D}N_j -\frac14 B'_{\mu\nu}B'^{\mu\nu} \nonumber \\ &&+\left( Y_{ij} L_i\mathcal{H} N_j +\frac12 y_{ij} A N_iN_j + \mbox{h.c.}\right)   \nonumber \\ &&-\lambda_a|A|^4+\lambda_{ah} |A|^2|\mathcal{H}|^2,
\eea 
where $\Lag^{\rm ns}_{\rm SM}$  represents the classically scale-invariant SM Lagrangian with gauged $B-L$ and the $L_i$ are the three families of SM lepton doublets.  
Here $D_\mu$ is the covariant derivative with respect to the full gauge group $SU(3)_C\times SU(2)_L\times U(1)_Y\times U(1)_{B-L}$:
\bea D_\mu &&= \partial_\mu +i g_3 T^\alpha G^\alpha_\mu + i g_2 T^a W^a_\mu + i g_Y {\mathcal Y} B_\mu  \nonumber \\ && + i \left[g_m {\mathcal Y}+ g_1' (B-L)\right] B'_\mu,\label{CovBmL}\eea 
which involve the gluons $G^\alpha_\mu$, the triplet of $W$ bosons $W^a_\mu$ as well as the gauge fields $B_\mu$ and $B_\mu'$ of $U(1)_Y$ and $U(1)_{B-L}$ (as usual $B'_{\mu\nu} \equiv \partial_\mu B'_\nu-\partial_\nu B'_\mu$) together with the respective generators $T^\alpha, T^a, {\mathcal Y}, B-L$ and gauge couplings $g_3, g_2, g_Y, g_1'$. Here $g_m$ corresponds to the Abelian mixing between $U(1)_Y$ and $U(1)_{B-L}$. 
This CSI matter Lagrangian has been previously considered in Refs.~\cite{Iso:2009ss,Salvio:2023ynn}, but without accounting for the observational signals of new physics and inflation at the same time.

Including the dynamics of gravity, the  Lagrangian of the model is given by
\be \Lag =\Lag_{\rm matter}-\frac12 f(\phi) R-\Lambda_0, \ee
where 
\be f(\phi) \equiv  \bp^2+ 2\xi_a |A|^2+2\xi_h|\mathcal{H}|^2\ee
includes the non-minimal couplings between the scalars $\phi$ and the Ricci scalar $R$~\cite{phi}.
 In the presence of gravity the action is $S = \int d^4x \sqrt{|\det \hat g|} \Lag$, where  $\det \hat g$ is the determinant of the spacetime metric.
The non-minimal couplings $\xi_a$ and $\xi_h$ respect classical scale invariance. In this work we add the standard Einstein-Hilbert term with the reduced Planck mass $\bp$ and a cosmological constant term $\Lambda_0$, which explicitly break classical scale invariance. However, it is possible to introduce another sector of the theory that can also generate $\bp$ and $\Lambda_0$ radiatively~\cite{Salvio:2014soa,Kannike:2015apa,Salvio:2017qkx,Salvio:2020axm,Alvarez-Luna:2022hka,Cecchini:2024xoq}. The specific choice and analysis of such a sector is beyond the scope of this paper and is left for future work. It is always possible to choose the typical energy scales of this sector large enough that the  results discussed in this paper are valid. 
In this context, it may be possible to remove, or dynamically generate, another hidden fundamental mass scale (the cutoff) that appears even at the classical level~\cite{Salvio:2017qkx,Anselmi:2017ygm,Anselmi:2018ibi,Donoghue:2019fcb,Salvio:2024joi,Salvio:2022mld}, a necessary step to obtain a fully CSI theory. Here we do not attempt to achieve this ambitious goal and to give an explanation of the observed value of the cosmological constant.

The one-loop quantum effective potential renormalized at $\tilde\mu$ and along the flat direction has the well-known Coleman-Weinberg form~\cite{Coleman:1973jx,Gildener:1976ih}
\be V_q(\chi) = \frac{\bar \beta}4\left(\log\frac{\chi}{\chi_0}-\frac14\right)\chi^4,\label{CWpot}\ee
where $\bar \beta$ is the beta function of the quartic coupling that vanishes at $\tilde\mu$ (see the appendix for the relevant one-loop beta functions in this model).   In our setup
\begin{equation}
    \bar \beta\approx \frac{96 g'^4_{1}}{(4\pi)^2}>0. \label{bbetag1p}
\end{equation}
 Also $\chi_0$ is the VEV of $\chi$.

Since $\chi_0\neq0$ breaks $U(1)_{B-L}$ the spin-1 particle $Z'$ associated with $U(1)_{B-L}$ acquires a mass $m_{Z'}\equiv m_{Z'}(\chi_0)$, where
\be m_{Z'}(\chi) = 2 g_1' \chi \label{bdm}\ee
is the background-dependent mass (we take $g_1' >0$). Also the fluctuation field  $\chi-\chi_0$ acquires a mass $m_\chi=\sqrt{\bar\beta}\chi_0$.
If $\lambda_{ah}\neq0$ this RSB also induces EW symmetry breaking.

The expression in~(\ref{CWpot}) is the universal form of the one-loop quantum effective potential. However, the relation between $\chi_0$ and $\tilde \mu$ is renormalization-scheme dependent: the definition of renormalized couplings is given in some renormalization scheme. Let us use the modified minimal subtraction ($\overline{\rm MS}$) scheme. A simple relation between $\chi_0$ and $\tilde \mu$ can be found noting that the largest contribution to $V_q$ in our setup is given by the $U(1)_{B-L}$ gauge vector so in the $\overline{\rm MS}$ scheme we get 
 \begin{equation}
 	V_q(\chi) \approx \frac{3}{64 \pi^2}m_{Z'}^4(\chi)\left[\log\left(\frac{m^2_{Z'}(\chi)}{\tilde \mu^2}\right)-\frac{5}{6}\right].
 \end{equation}
Eq.~(\ref{bdm}) then leads to
\begin{equation}
	\label{eq:phimut}
	\chi_0 \approx \tilde \mu \frac{e^{1/6}}{2g_1'}, \hspace{0.4cm} m_\chi \approx \frac{\sqrt{6}}{2\pi}e^{1/6}g_1' \tilde \mu,\hspace{0.4cm} m_{Z'}\approx e^{1/6}\tilde \mu. \nonumber
\end{equation}

The quantum corrections to the derivative part of the effective action can be neglected for our purposes: essentially we will use the effective potential during inflation  and to study the FOPT. During inflation the temperature is absent and the quantum corrections to the derivative part are suppressed as usual by  loop factors, $\sim 1/(4\pi)^2$. In the FOPT supercooling guarantees that those quantum corrections are small even if $T\neq 0$~\cite{Salvio:2023qgb}.

\section{Inflation}\label{Inflation}

The action can be simplified by performing a Weyl transformation (sometimes called ``conformal transformation") of the metric 
\be g_{\mu\nu} \rightarrow \frac{\bp^2}{f}g_{\mu\nu}. \label{ConfTransf}\ee
The resulting field variables constitute what is known as the Einstein frame, while the original field variables form the Jordan frame. 
After this Weyl transformation the scalar-tensor part of the Lagrangian is given by
\be \Lag_{st} = \bp^2 \left[\frac{\partial_\mu \phi\partial^\mu \phi}{2f} +\frac{3\partial_\mu f\partial^\mu f}{4f^2}\right]- U -\frac{\bp^2}{2} R, \label{SstTransf}\ee
where $U = \bp^4V/f^2$
and $V$ is the effective potential in the Jordan frame.

Here we explore the possibility that inflation is driven by  $\chi$. In general two scalars, $A$ and $\mathcal{H}$, can take part in inflation. However, by setting $\xi_h$ negative and large enough it is possible to render $\mathcal{H}$ inactive during inflation and the metastability of EW vacuum~\cite{Buttazzo:2013uya} harmless at the same time~\cite{Joti:2017fwe} (see also Refs.~\cite{Rajantie:2017ajw,Markkanen:2018pdo}). The phase of $A$, on the other hand, provides the longitudinal component of $Z'$ as $\chi_0\neq 0$ and $U(1)_{B-L}$ is spontaneously broken. In this setup $f=\bp^2 + \xi \chi^2$, for an appropriately defined coupling $\xi$, and $V$ coincides with $V_q$ in~(\ref{CWpot}). Moreover, we can introduce a canonically normalized scalar $\chi_E$ (which is a function of $\chi$). 

 \begin{figure}[t!]
\begin{center}
 \includegraphics[scale=0.56]{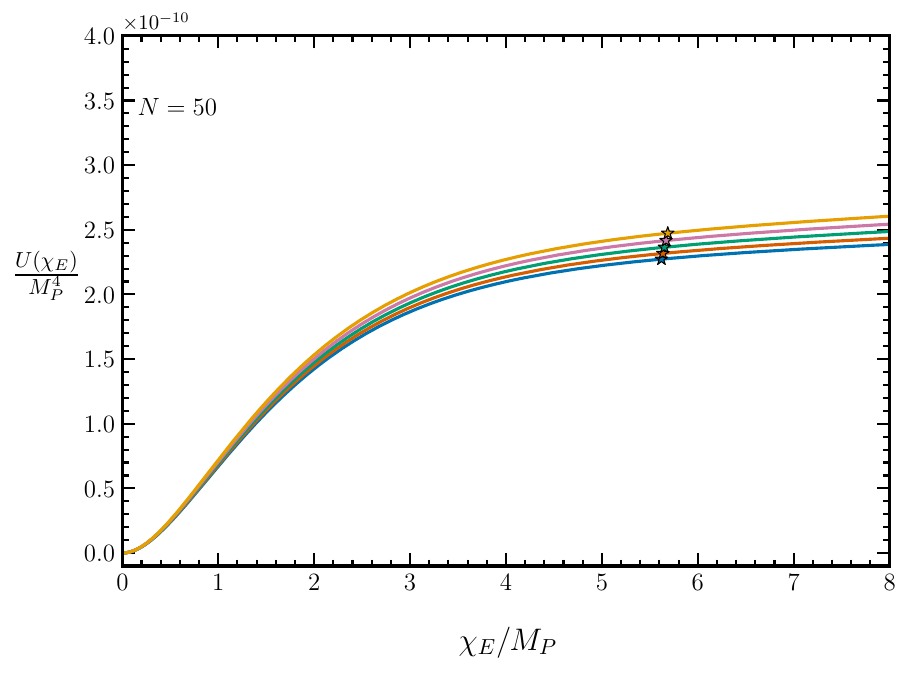} \\
  \includegraphics[scale=0.56]{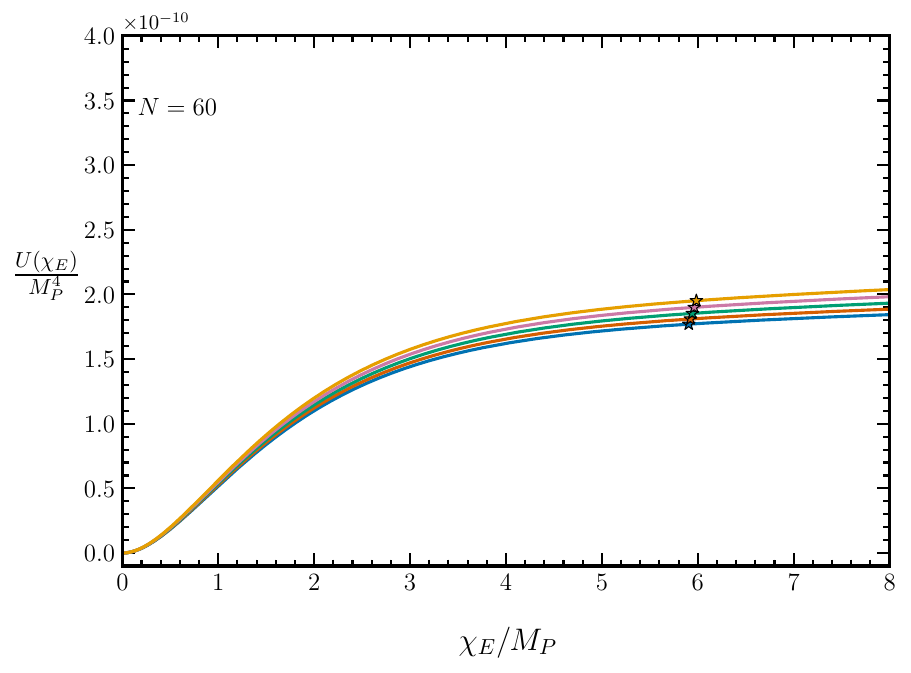}
 \end{center} 
 \vspace{-.5cm}
   \caption {\em Einstein-frame potential. The stars correspond to the field values $N$ e-folds before the end of inflation and the  following benchmark points are considered:} 
   \begin{itemize}
   \item blue ($\bar\beta\approx0.0028, ~\chi_0\approx9.7\times 10^3$\,GeV),
    \item orange ($\bar\beta\approx0.0032, ~\chi_0\approx2.8\times 10^4$\,GeV), 
    \item green ($\bar\beta\approx0.0036, ~\chi_0\approx8.3\times 10^4$\,GeV), 
    \item pink ($\bar\beta\approx0.0038, ~\chi_0\approx2.4\times 10^5$\,GeV), \item yellow ($\bar\beta\approx0.0040, ~\chi_0\approx7.1\times 10^5$\,GeV).    \end{itemize}
\label{potential}
\end{figure}

 \begin{figure}[t!]
\begin{center}
 \includegraphics[scale=0.56]{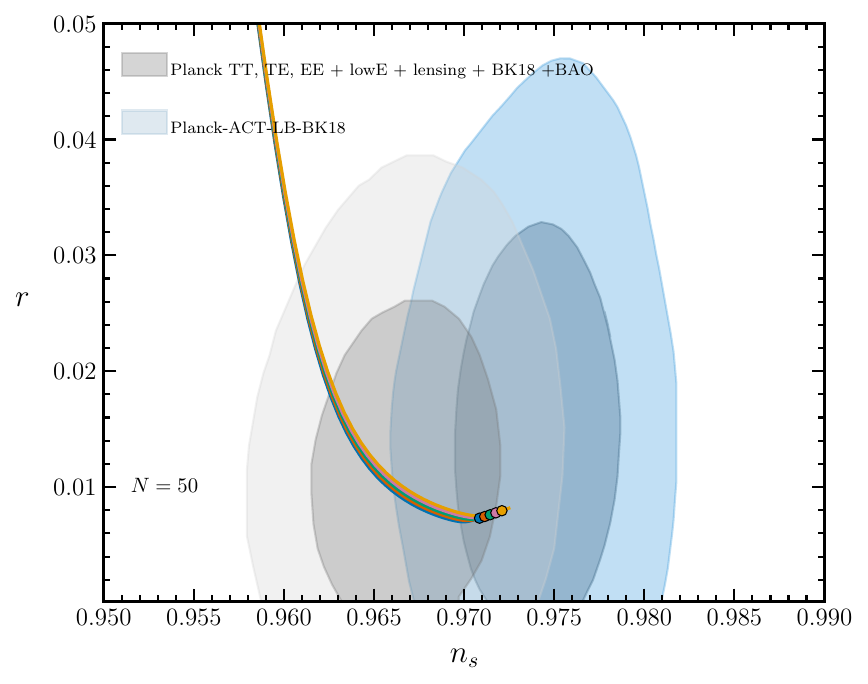} \\
\hspace{-0.3cm}\includegraphics[scale=0.56]{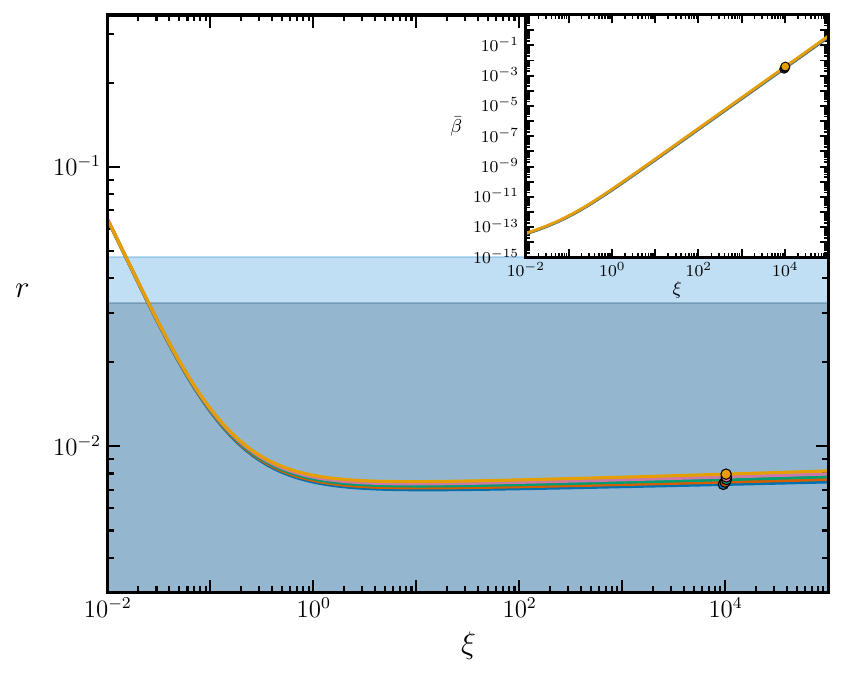}
 \end{center} 
 \vspace{-.5cm}
   \caption{\em 
   Predictions of the model for $n_s$ and $r$ together with the bounds from~\cite{Ade:2015lrj,BICEP:2021xfz} (in gray) and  the more recent ones from~\cite{ACT:2025fju,ACT:2025tim}. The dot corresponds to the benchmark point given in the caption of Fig.~\ref{potential} with the same color. The other points on the line are obtained by taking different values of $\bar\beta$ (and thus of $\xi$) as indicated in the inset, but keeping the same value of $\chi_0$. Here $N=50$.
   }
\label{nsr}
\end{figure}

 \begin{figure}[t!]
\begin{center}
 \includegraphics[scale=0.56]{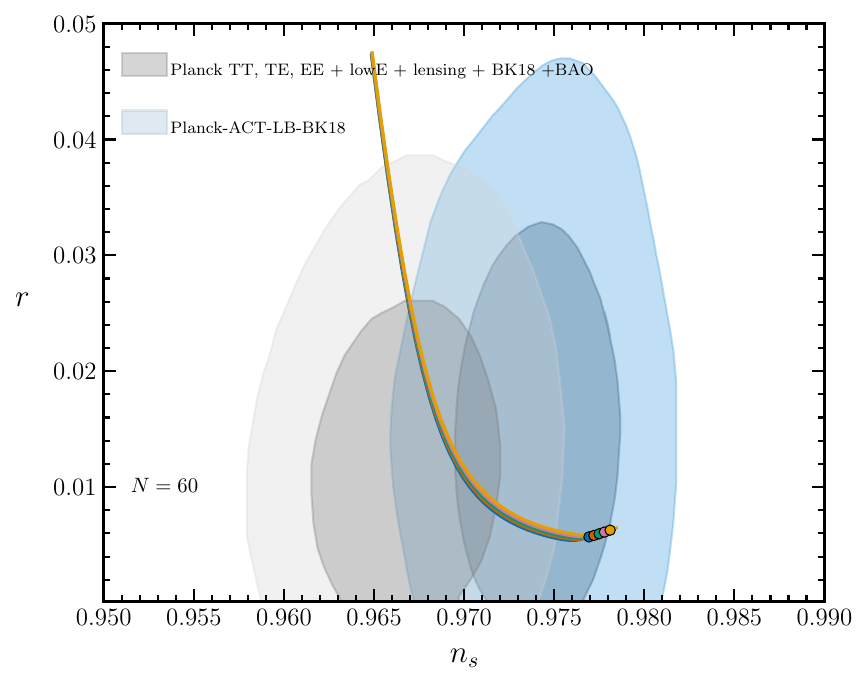} \\
 \hspace{-0.3cm} \includegraphics[scale=0.56]{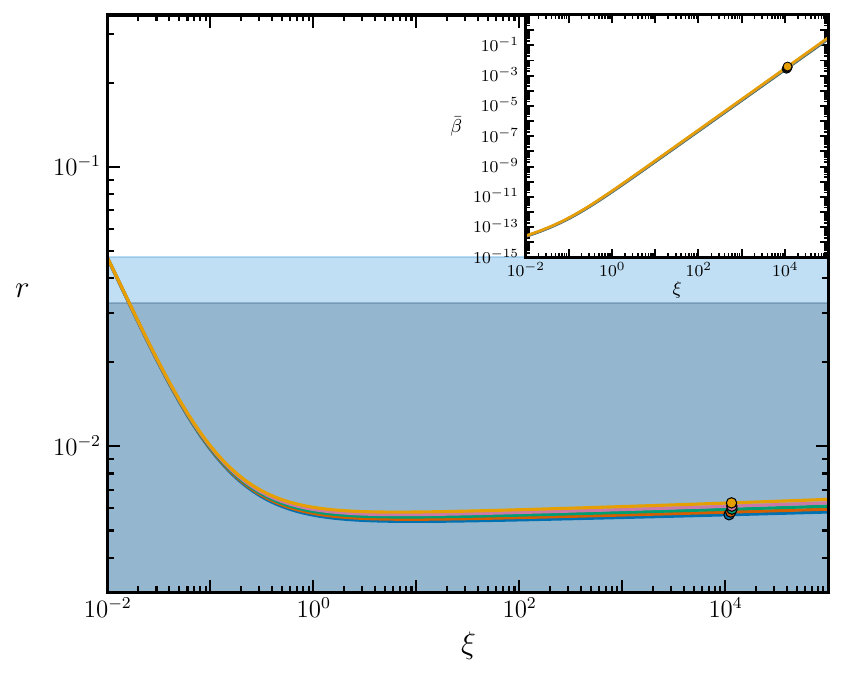}
 \end{center} 
 \vspace{-.5cm}
   \caption{\em 
   The same as in Fig.~\ref{nsr} but for $N=60$.
   }
\label{nsr60}
\end{figure}

In the slow-roll approximation we can use well-known expressions to compute the number of e-folds $N$ during inflation,   $n_s$, the tensor-to-scalar ratio $r$ and the curvature power spectrum $P_R$ (at horizon exit). Imposing the observed value of $P_R$ (we use $(2.10 \pm 0.03) \times 10^{-9}$ given in~\cite{Ade:2015lrj}) one obtains $\bar\beta$ as a function of $\xi$.

Figs.~\ref{potential} and~\ref{nsr}-\ref{nsr60} show respectively $U(\chi_E)$ and the predictions of the model for $n_s$ and $r$ for some benchmark points and $N=50,60$. We checked that the slow-roll approximation used is valid. Our results provide a realization of the inflationary results of~\cite{Gialamas:2025kef} within a concrete and realistic model. In Figs.~\ref{nsr}-\ref{nsr60} we also show the constraints of the  ACT collaboration~\cite{ACT:2025tim} and the Planck and BK18 collaborations~\cite{BICEP:2021xfz}. As explained above, our setup effectively leads to a single-field inflationary model, which satisfies the assumptions of Refs.~\cite{ACT:2025tim,BICEP:2021xfz}. The spectrum of primordial fluctuations is almost, but not quite, scale invariant; non-adiabatic perturbations and non-Gaussianities are negligibly small.

One interesting feature of the predictions in Figs.~\ref{nsr}-\ref{nsr60} is the presence of a minimal value of $r$, which is within the reach of future CMB observations, such as those of LiteBIRD~\cite{LiteBIRD:2022cnt} and CMB-S4~\cite{CMB-S4:2016ple}. This minimal value depends very weakly on $\chi_0$ for our benchmark points (given in the caption of Fig.~\ref{potential}). For reasons that will be clear in Sec.~\ref{Reheating and thermal inflation}, taking $\chi_0$ much below the minimal one in the benchmark points would lead to a scalar $\chi-\chi_0$ with a mass $m_\chi$ too small  and a coupling $\lambda_{ah}$ too large to be phenomenologically viable. On the other hand,  taking $\chi_0$ much above the maximal one in the benchmark points, would lead to extremely tiny values of $\lambda_{ah}$, which we want to avoid. As a result, in the region of parameter space we are interested in, the minimal value of $r$ is very close to that in Figs.~\ref{nsr}-\ref{nsr60}.

The benchmark points are chosen in a way that $\bar\beta$ (and thus $g_1'$, cf.~Eq.~(\ref{bbetag1p})) is not too small because of the reasons discussed in Sec.~\ref{Building the model}. We will explain in more detail the motivations for this setup in Sec.~\ref{Reheating and thermal inflation}. Now we can note, looking at Figs.~\ref{nsr}-\ref{nsr60}, that these benchmark points correspond to a rather large value of $\xi$ (of order $10^4$). In the context of Higgs inflation~\cite{Cervantes-Cota:1995ehs,Bezrukov:2007ep} such a large value of $\xi$ leads to a breaking of perturbative unitarity in flat spacetime at the energy scale $\bp/\xi$~\cite{Burgess:2009ea,Barbon:2009ya,Burgess:2010zq,Hertzberg:2010dc,Burgess:2014lza} when the inflaton is at the point of minimum of the potential, which is much smaller than $\bp/\xi$. However, as pointed out in~\cite{Bezrukov:2010jz}, the cutoff of the theory is significantly larger than $\bp/\xi$ when the inflaton is much larger than $\bp/\xi$, which is the case during inflation, such that a value of $\xi$ of order $10^4$ is still within the regime of validity of the effective theory. Although in our case the inflaton potential is different than in Higgs inflation, we have checked that the same conclusion holds in our model too. 
 
It is important to note that $N$ here is the number of e-folds during slow-roll inflation, but it is not the full number of inflationary e-folds $N_e$ because a period of thermal inflation is predicted by the present model after a radiation-dominated era, which follows the initial (slow-roll) inflation, as we will discuss in Sec.~\ref{Reheating and thermal inflation}.


\section{Reheating and thermal inflation}\label{Reheating and thermal inflation}

Just after the initial phase of inflation we have described in Sec.~\ref{Inflation},  reheating takes place. If the portal coupling $\lambda_{ah}$ is not extremely small the reheating temperature $T_{\rm rh}$ is much above $\chi_0$. This occurs because $\chi$ has a non-vanishing (although small) component along $\mathcal{H}$. 

To illustrate this point note that in any gauge the classical scalar potential in the Jordan frame is 
\begin{equation}
    \label{eq:VB-L}
     V_{\rm cl}\left(h,a\right)=\frac{1}{4}\lambda_{a} a^4+\frac{1}{4}\lambda_{h} h^4-\frac{1}{4}\lambda_{ah}h^2a^2,
\end{equation}
where $\lambda_h$ is the Higgs quartic coupling.
We have decomposed $\mathcal{H}$ and $A$ into real scalar fields $h$, $\pi_i$, $a$ and $\pi_a$:
\begin{equation}
    \mathcal{H}=e^{i\pi_i\tau_i}  \left(\bac 0 \\
\frac{h}{\sqrt{2}} \ea \right), \qquad 
    A\equiv \frac{a}{\sqrt{2}} e^{i\pi_a}, \label{ahDef}
\end{equation}
where the $\tau_i$ are the three Pauli matrices, 
such that in any gauge the potential involves only two real gauge-invariant fields, $h$ and $a$. One can express them in terms of $\chi$ and the (observed) Higgs boson field $H$:
\begin{equation}
    \begin{cases}\label{eq:rotation}
    a=\chi \cos \alpha- H\sin \alpha,\\ 
    h=\chi \sin \alpha+H \cos \alpha,
    \end{cases}
\end{equation}
where the mixing angle $\alpha$ is defined by
\begin{equation}
    \tan \alpha\equiv\sqrt{\frac{\lambda_{a h}}{2\lambda_h}}. \label{alphaDef}
\end{equation}
So $\lambda_{\phi h}\neq0$ ensures that $\chi$ has a non-vanishing component along $h$. This leads to a portal between the inflationary and the SM sectors, which allows EW RSB and reheating. 

To show that $T_{\rm rh}$ is much above $\chi_0$ for some viable values of $\lambda_{ah}$,  note that after inflation ends the inflaton performs large oscillations around the point of  minimum of the potential. In particular there are regions of the potential where the effective inflaton mass,
\begin{equation}
	 \frac{d^2\, U}{d\,\chi_E^2} = \left(\frac{d\,\chi_E}{d\,\chi}\right)^{-1}\frac{d}{d\chi} \left(\left(\frac{d\,\chi_E}{d\,\chi}\right)^{-1}\frac{d\,U(\phi(\chi))}{d\,\chi}\right),
\end{equation}
  is much larger than the EW scale. We call $M_{\rm eff}^2$ the maximal value of the  effective inflaton mass with respect to $\chi$, which is assumed for some value, $\chi_{\rm max}$, of $\chi$. When the inflaton performs these large oscillations it decays into SM particles and reheats the universe up to 
  \begin{equation}
	T_{\rm rh}\approx\left(\frac{45\Gamma^2_{\chi_E}\bp^2}{4\pi^3g_*(T_{\rm rh})}\right)^{1/4},
\end{equation}  
where the fast-reheating approximation is used. Here $\Gamma_{\chi_E}$ is the total decay width of $\chi_E$ and $g_*(T)$ is the effective number of relativistic species at temperature $T$. One of the main decay channels is that into a top-antitop pair, with decay width 
  \be \Gamma_{\chi_E\to t\bar t} \approx \left(\frac{d\,\chi}{d\,\chi_E}\right)_{\chi=\chi_{\rm max}}^{2} \hspace{-0.3cm}\frac{3y_t^2 \sin^2\alpha}{16\pi} M_{\rm eff},\ee 
  where $y_t$ is the top Yukawa coupling. For the benchmark points given  in the caption of Fig.~\ref{potential}, which are representative of the setup adopted in this work, this channel alone reheats the universe up to a temperature at least of order $10^{10}$\,GeV$\gg\chi_0$.

A precise calculation of $T_{\rm rh}$ is an interesting target for future research. However, it is not needed here: all we will need to know regarding $T_{\rm rh}$ in the rest of the paper is that  $T_{\rm rh}\gg\chi_0$ and, therefore, $T_{\rm rh}$ significantly exceeds the critical temperature above which RSB does not occur.

As a result, after reheating, $\chi$ is trapped in the symmetry preserving vacuum, $\chi=0$. When the temperature becomes small enough the universe undergoes a phase of thermal inflation with Hubble rate 
\be H_t=\frac{\sqrt{\bar\beta} \chi_0^2}{4\sqrt{3}\bp}.\ee
We note that this thermal inflation is realized with a potential similar to that used in warm inflation in~\cite{Berera:1995ie}.
 We leave for future work to establish whether during such thermal inflation a significant amount of particles are produced too. Understanding this point is interesting, but not essential for this work. 

During thermal inflation $T$ decreases exponentially (supercooling) and the effective potential increasingly resembles the zero-temperature one, Eq.~(\ref{CWpot}), except for a barrier that  obstructs the decay of the false vacuum $\chi=0$ and is always present as long as $T\neq0$~\cite{Salvio:2023qgb}. Note that supercooling guarantees that higher-loop corrections are suppressed by loop factors of the same order as the zero-temperature ones, $\sim1/(4\pi)^2$~\cite{Salvio:2023qgb}. Therefore, the relative uncertainties due to neglecting higher-loop corrections is of order $1/(4\pi)^2$. Note also that, using the {\it gauge-invariant} scalar fields $h$ and $a$ defined in Eq.~\eqref{ahDef}, the corresponding thermal effective action for $\chi$ is gauge independent because the effective action inherits the linear symmetries of the action~\cite{Weinberg:1996kr} and $\chi$ is gauge independent, see Eq.~\eqref{eq:rotation}. The bubbles of the true vacuum are diluted by the expansion of the universe as long as the decay rate per unit of volume, $\Gamma_{\rm v}$, is much smaller than $H_t^4$. When $T$ is small enough the condition 
\be \Gamma_{\rm v}\approx H_t^4 \label{TnEq}\ee
 might be realized and the PT starts being effective. The value of $T$ such that Eq.~(\ref{TnEq}) is satisfied (if any) is called nucleation temperature, $T_n$. 

If there is a solution of Eq.~(\ref{TnEq}) for $T_n$ the universe undergoes a nearly-exponential expansion from roughly  the time when the radiation energy density equals the vacuum energy density with temperature 
\be T_{\rm eq} \equiv \left(\frac{15\bar{\beta}\chi_{0}^4}{8\pi^2 g_{*}(T_{\text{eq}})}\right)^{1/4} \ee 
 and the nucleation time with temperature $T_n$. The corresponding number of e-folds can then be estimated as 
\be N_t\approx \log\left(\frac{T_{\text{eq}}}{T_n}\right). \ee

There are two possible situations. Either the spacetime curvature is irrelevant in the vacuum decay process or not. The former (latter) case occurs when $T_n$ is (not) much larger than the typical curvature scales.
%
More precisely, the spacetime curvature remains negligible  as long as the characteristic size $R_b$ of the bounce is much smaller than the linear size $\pi /H_t$ of the Euclidean de Sitter spacetime and also the effective mass of $\chi$ receives a negligible contribution from the non-minimal coupling $\xi$. In general $R_b\sim 1/m$, where $m\equiv g T/\sqrt{12}$ and $g$ plays the role of a ``collective coupling" of $\chi$ with all fields of the theory~\cite{Salvio:2023qgb,Salvio:2023ynn}: in our setup $g\approx2\sqrt{3} g_1'$. So the condition for the spacetime curvature to be negligible in the vacuum decay process is 
\be m\gg\max\left(\sqrt{12 |\xi|},\frac1{\pi}\right) H_t. \ee

When the temperature becomes so small that this condition is not satisfied anymore one should take into account the spacetime curvature. If $\xi\gg 1/(12\pi^2)$ the spacetime curvature manifests itself just as a negative correction to the effective squared mass of $\chi$, which favors and eventually always leads to the completion of the transition even if a solution of  Eq.~(\ref{TnEq}) for $T_n$ does not exist~\cite{Salvio:2023blb}. The number of e-folds $N_t$ during thermal inflation is then approximately given by
\be N_t \approx \log\left(\frac{gT_{\rm eq}}{12 \sqrt{\xi} H_t}\right). \label{NtApp}\ee
Setting $\chi_0$ not too many orders of magnitude above the EW scale (to avoid extremely small values of $\lambda_{ah}$) and avoiding to choose $\xi$ extremely large, $N_t$ turns out to be quite large, around some tens. Taking $\xi$ not much larger than $1/(12\pi^2)$ leads to an even larger $N_t$ because the right-hand side of~(\ref{NtApp}) is a decreasing function of $\xi$. Setting $\xi<0$ leads to an even larger $N_t$ because a negative $\xi$ impede the transition. This situation renders problematic the realization of inflation with an overall number of inflationary e-folds $N_e$ around 60.  For this reason we restrict the analysis to values  of the parameters such that the spacetime curvature is irrelevant in the vacuum decay process.

In this case we can use the model-independent formulas of~\cite{Salvio:2023ynn}, which are valid for large-enough supercooling: the quantity 
\be \epsilon\equiv  \frac{g^4}{6\bar\beta \log\frac{\chi_0}{T}}
 \label{CondConv}\ee
 should be below 10 in order of magnitude for the relevant temperatures, $T\approx T_n$.
 
    \begin{figure}[t!!]
 \hspace{-0.5301cm}\includegraphics[scale=0.44]{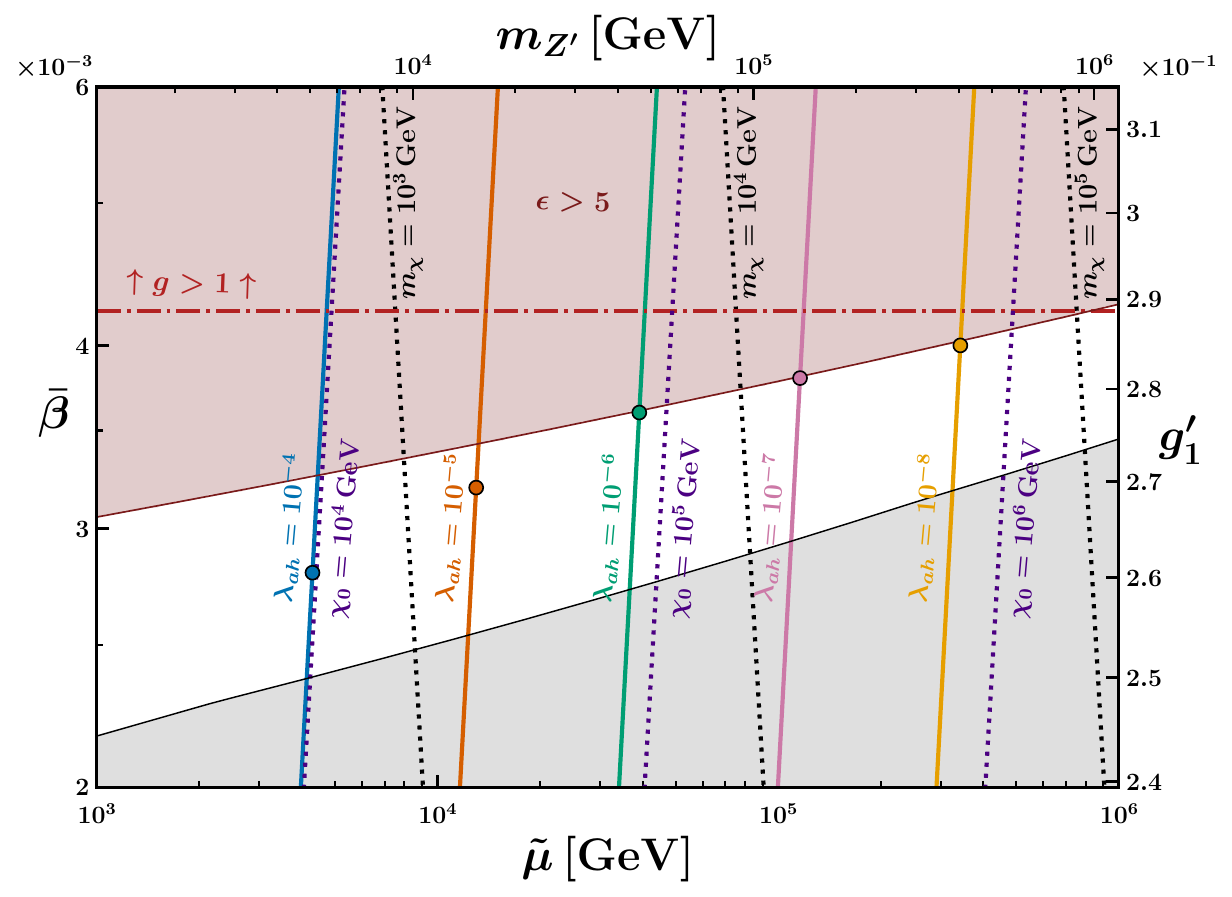}
 \vspace{-.5cm}
   \caption{\em Values of $\lambda_{ah}$, $\chi_0$ and $m_\chi$ as functions of $m_{Z'}$ and $g_1'$. The dots represent the benchmark points in the caption of Fig.~\ref{potential} and in Table~\ref{table}.
   }
   \label{parameters}
\end{figure}

 The results in our setup are summarized in Fig.~\ref{parameters} paying particular attention to the benchmark points in the caption of Fig.~\ref{potential} and Table~\ref{table}, which represent some examples of our viable setup.
 \begin{table}[h!]
\centering
\begin{tabular}{|c|c|c|}
\hline
\textbf{BM} & $T_n$ [GeV] & $N_t$ \\
\hline
blue & 0.35 & 7 \\
orange &  4.1& 6 \\
green  & 30& 5  \\
pink & 88& 5 \\
yellow &  $2.4\times10^2$ & 5 \\
\hline
\end{tabular}
\caption{\it Approximate values of the nucleation temperature $T_n$ and the number of e-folds $N_t$ during thermal inflation for the benchmark points (BMs) given in the caption of Fig.~\ref{potential}. 
}
\label{table}
\end{table}
%

At the scale $\tilde \mu$ where the flat direction occurs we imposed the conditions 
\begin{equation}
	\lambda_{ah}=2\sqrt{\lambda_{a}\lambda_{h}}, \hspace{1cm} v=\sin \alpha\, \chi_0,
\end{equation}
where $v$ is the SM VEV. Using~(\ref{alphaDef}), these equations allow us to determine $\lambda_{ah}$ and $\lambda_a$ at the scale $\tilde \mu$ in terms of $\chi_0$, which are used in Fig.~\ref{parameters}. The gauge mixing  parameter $g_m$ has been fixed to a negligibly small value at the scale $\tilde\mu$.
We observe that the smaller $\chi_0$ is the bigger $\lambda_{ah}$ turns out to be, as shown in Fig.~\ref{parameters}. So extremely small values of $\lambda_{ah}$ can be avoided  when the mass and coupling of the new degree of freedom $\chi$ required by RSB get closer to the reach of the Large Hadron Collider. All benchmark points shown in Fig.~\ref{parameters} are, however, all well within the experimental bounds. 

The GW spectrum and the primordial-black-hole abundance, mass and spin that are predicted by this RSB model can be extracted from the model-independent results of~\cite{Salvio:2023ynn,Banerjee:2024cwv}. For example, considering the benchmark points in Fig.~\ref{parameters}, the predicted GWs can be tested by the future Laser Interferometer Space Antenna (LISA)~\cite{Babak:2021mhe} and the fraction of PBH dark matter produced is very small, less than $\sim1\%$ for
$\chi_0 \sim10^5$\,GeV.

Reheating after thermal inflation in this model is guaranteed by the sizable couplings of the inflaton to the SM particles and produces sterile neutrinos, which can account for the observed dark matter abundance~\cite{Rescigno:2025ong}.

\begin{figure}[t!!]
 \hspace{-0.5cm}\includegraphics[scale=0.57]{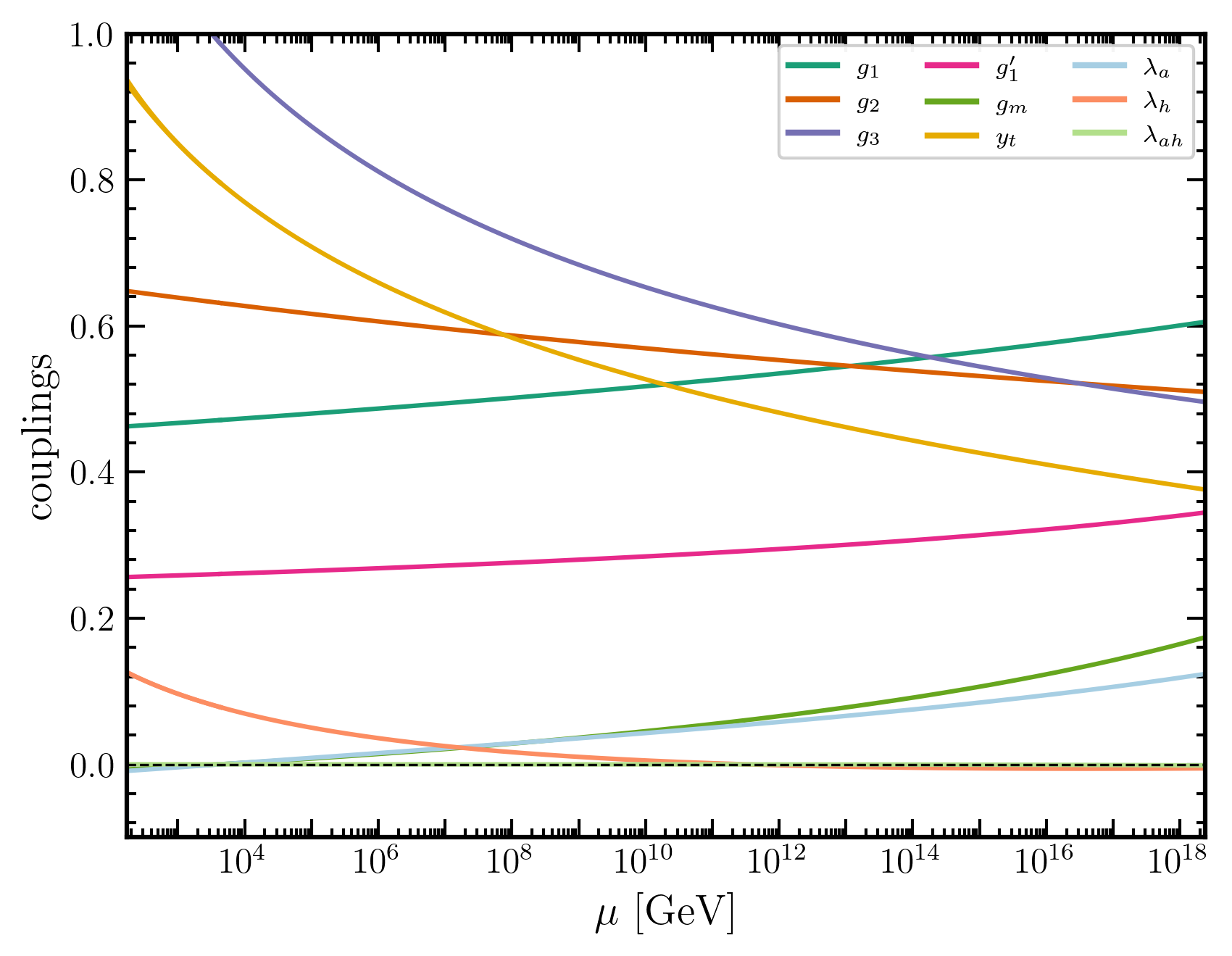}
 \vspace{-.5cm}
   \caption{\em Running of the couplings. 
   The initial conditions correspond to the blue benchmark point. 
   All Yukawa couplings different from $y_t$ are so small to be invisible in this plot for our setup.
   }
   \label{Running}
\end{figure}

We also verified, by employing the renormalization group equations (RGEs) given in the appendix, that the benchmark points in Fig.~\ref{parameters} are not affected by Landau poles and preserve perturbativity up to the Planck scale. In Fig.~\ref{Running} we show the running couplings corresponding to the blue benchmark point. Figures for  other benchmark points are given in the appendix. One aspect that is perhaps worth noting is that the smallness of $\lambda_{ah}$ is preserved by the RG evolution.

\section{Conclusions}\label{Conclusions}

 We constructed and explored a fully realistic and well-motivated CSI model that accounts for all evidence of new physics and  is compatible with the most recent constraints on inflationary observables. Classical scale invariance offers a way to understand large mass hierarchies and  increases the predictivity of the model as masses are radiatively generated, they are not independent parameters (see e.g.~\cite{Ghoshal:2020vud}). 
  
  CSI perturbative theories suggest a non-standard cosmology in which inflation occurs in two stages, separated by a radiation-dominated era, as illustrated by our well-motivated realistic model. Therefore, this model can be considered a concrete implementation of a ``rollercoaster cosmology"~\cite{DAmico:2020euu}.

 The two stages consist here of an initial slow-roll inflation and a thermal inflation associated with the first-order PT. The presence of thermal inflation is a generic prediction of CSI theories because RSB is always associated with a first-order PT~\cite{Witten:1980ez,Salvio:2023qgb}. The fact that some e-folds, $N_t$, are produced by thermal inflation implies that the inflationary observables ($n_s$, $r$, $P_R$, etc.) can be computed at a correspondingly lower value of e-folds $N$ before the end of slow-roll inflation. 

Given that thermal inflation produces in our setup a number of e-folds $N_t$ somewhat between 5 and 10, cf.~Table~\ref{table}, Fig.~\ref{nsr} shows that our model is compatible at $1\sigma$ level with remarkably both the bounds in~\cite{Ade:2015lrj,BICEP:2021xfz} (Planck and BK18 collaborations) and the more recent ones in~\cite{ACT:2025fju,ACT:2025tim} (ACT collaboration) for an acceptable number of total inflationary e-folds $N_e$ between 55 and 60.
 
 \subsection*{Acknowledgments}
This work was partially supported by the Italian Ministry of University and Research (MUR) under the grant PNRR-M4C2-I1.1-PRIN 2022-PE2 Non-perturbative aspects of fundamental interactions, in the Standard Model and beyond F53D23001480006 funded by E.U. - NextGenerationEU.

\section*{Appendix}

In this appendix we provide the one-loop RGEs of the model, neglecting the small contributions of the Yukawa couplings involving SM fermions lighter than the top. We have derived these RGEs by using the general approach of~\cite{MV}.
\begin{figure}[t!!]
 \hspace{-0.5cm}\includegraphics[scale=0.57]{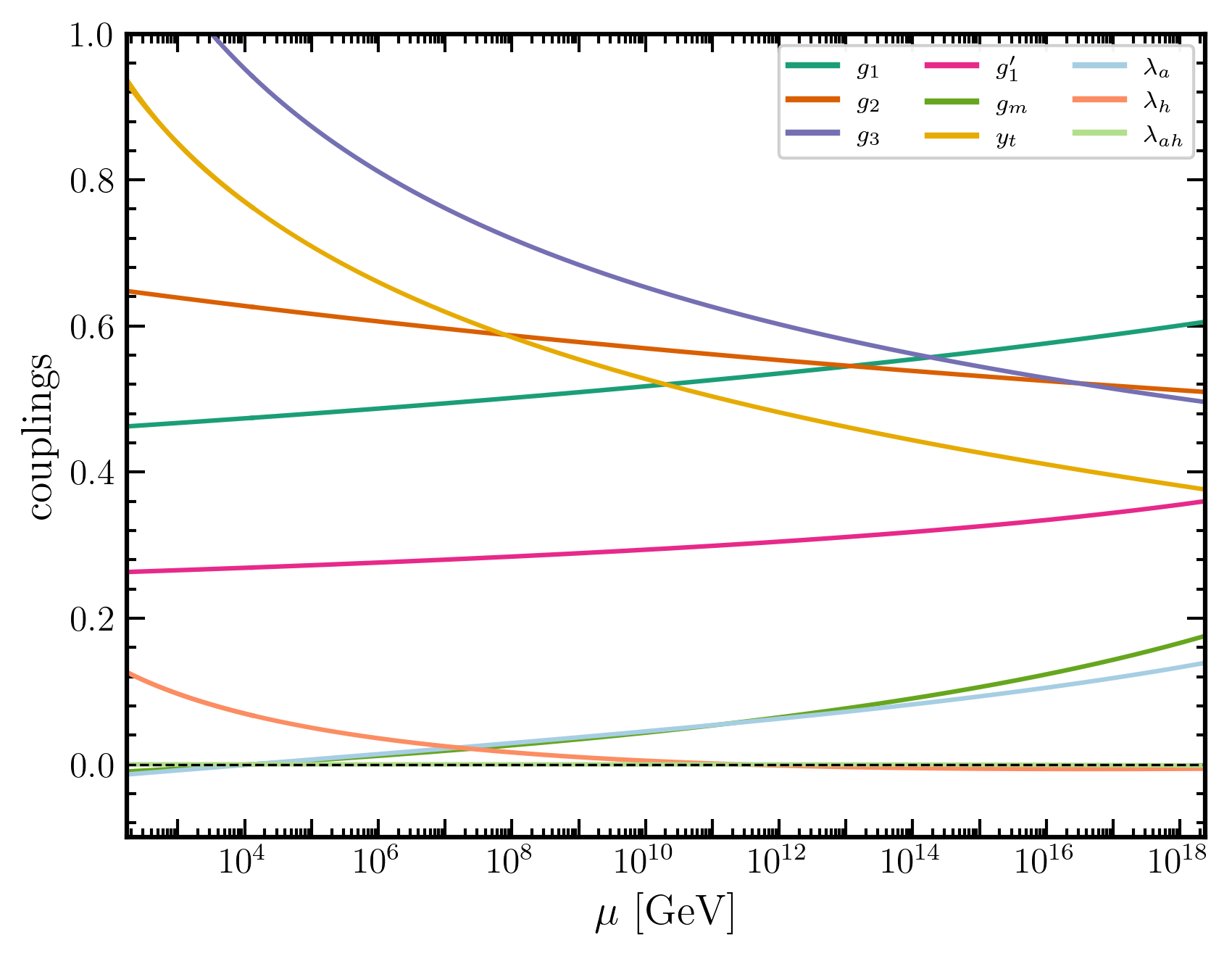}\\
\hspace{-0.3cm}\includegraphics[scale=0.57]{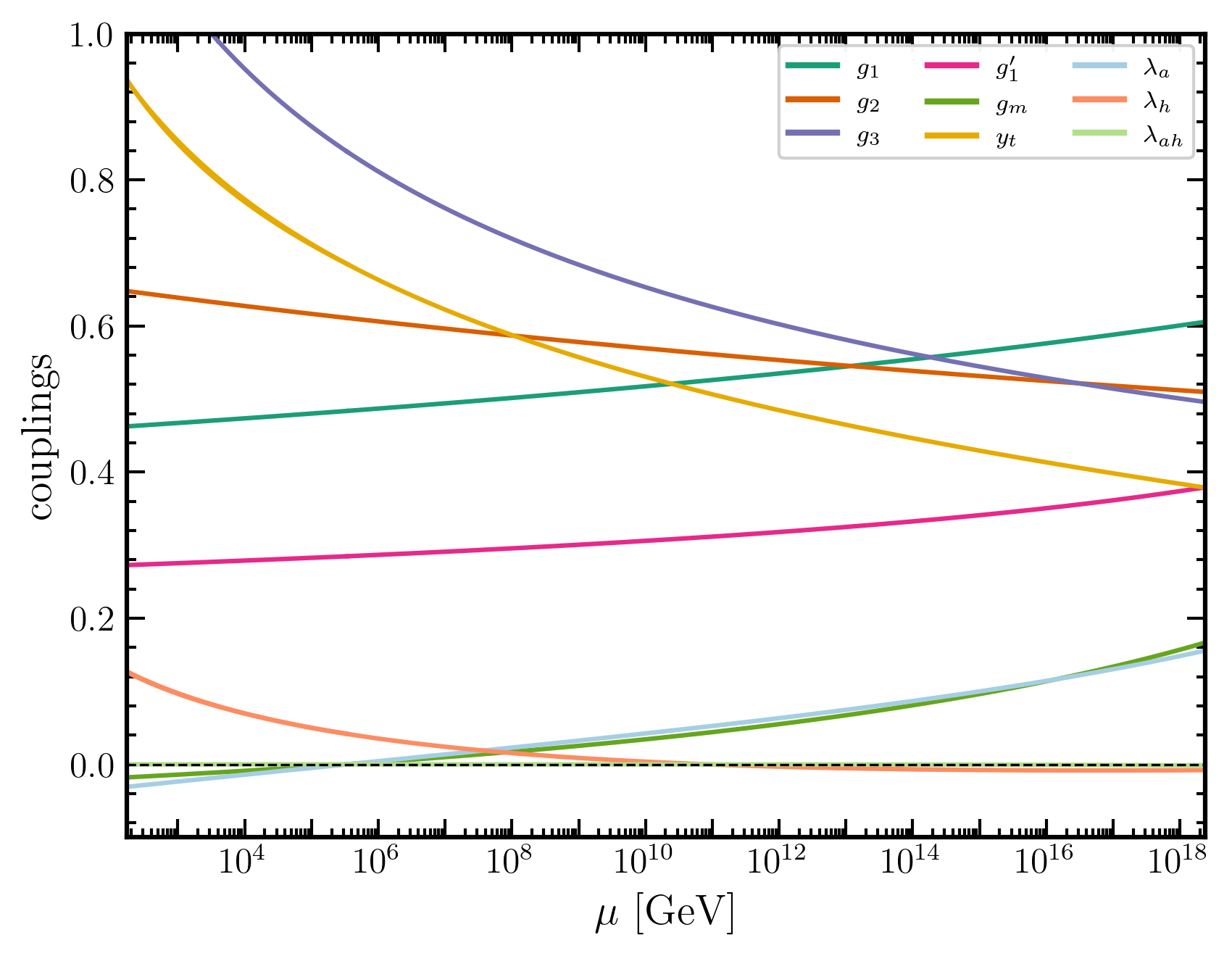}
 \vspace{-.5cm}
   \caption{\em Like in Fig.~\ref{Running} but for the orange (upper plot) and yellow (lower plot) benchmark points.
   }
   \label{RunningA}
\end{figure}
\bea
\frac{dg_1}{d \ln \mu}&=& \frac{g_1^3}{(4 \pi)^2}\frac{41}{10},\nonumber \\ [3pt]
    \frac{dg_2}{d \ln \mu} &=& \frac{g_2^3}{(4\pi)^2}\left(-\frac{19}{6}\right),\nonumber\\
    \frac{d g_3}{d\ln \mu} &=& \frac{g_3^3}{(4 \pi)^2} \left(-7\right),\nonumber \\
    \frac{dg_1'}{d \ln \mu} &=& \frac{1}{(4\pi)^2}\left(12g_1'^3+\frac{32}{3}g_1'^2g_m+\frac{41}{6}g_1'g_m^2\right),\nonumber\\
    \frac{dg_m}{d\ln \mu} &=& \frac{1}{(4\pi)^2}\left[\frac{41}{6}g_m\left(g_m^2+\frac{6g_1^2}{5}\right)+ \frac{32}{3}g_1'\left(g_m^2+\frac{3g_1^2}{5}\right)\right. \nonumber \\ &&\left.+12g_1'^2g_m\right],\nonumber\\
    \frac{dy_t}{d \ln \mu} &=& \frac{y_t}{(4\pi)^2}\left(\frac92y_t^2-8g_3^2-\frac{9g_2^2}{4}-\frac{17g_1^2}{20}-\frac{17}{12}g_m^2\right. \nonumber \\ &&\left.-\frac{2g_1'^2}{3}-\frac{5g_mg_1'}{3}\right),\nonumber \\
    \frac{dy_i}{d\ln\mu}&=& \frac{y_i}{(4\pi)^2}\left(y_i^2+\frac12\sum_jy_j^2-6g_1'^2\right),\nonumber\\
    \frac{d \lambda_{a}}{d \ln \mu} &= &\frac{1}{(4\pi)^2}(20\lambda_{a}^2 + 2\lambda_{ah}
    ^2-\sum_jy_j^4+96g_1'^4\nonumber \\&&-48\lambda_{a}g_1'^2+2\lambda_{a}\sum_jy_j^2),\nonumber \\
  \frac{d\lambda_{h}}{d \ln \mu}&=& \frac{1}{(4\pi)^2}\left[\frac{9 g_2^4}{8}+\frac{9g_1^2g_2^2}{20}+\frac{27g_1^4}{200}-6y_t^2+24\lambda_{h}^2\right.\nonumber \\&&+\lambda_{ah}^2+\frac{3g_2^2g_m^2}{4}+\frac{9g_1^2g_m^2}{20}+\frac{3g_m^4}{8}\nonumber \\&&\left.+ \lambda_{h}\left(12y_t^2-\frac{9g_1^2}{5}-9g_2^2-3g_m^2\right)\right],\nonumber\\
    \frac{d\lambda_{ah}}{d\ln \mu}&=&\frac{1}{(4\pi)^2}\bigg\{\lambda_{ah}\bigg[12\lambda_{h}+8\lambda_{a}-4\lambda_{ah}+6y_t^2\nonumber\\&&\left.- 
    \frac{3}{2}\left(3g_2^2+\frac{3g_1^2}{5}+g_m^2\right)+\sum_jy_j^2-24g_1'^2\right]\nonumber\\&&-12g_m^2g_1'^2\bigg\},\nonumber
    \eea
 where $g_1\equiv \sqrt{5/3}g_Y$.
To simplify the expressions above we have put, without loss of generality, the Yukawa matrix with elements $y_{ij}$ in diagonal form with diagonal elements $y_i$ real by means of a unitary transformation acting on the $N_i$. This is known as the complex Autonne-Takagi factorization, see e.g.~\cite{Youla}.

In Fig.~\ref{RunningA} we show the running couplings corresponding to the orange and yellow benchmark point in Fig.~\ref{parameters}, obtained by solving the RGEs above.


\begin{thebibliography}{9}
\bibitem{indices}Greek indices are lowered and raised by the metric $g_{\mu\nu}$, for which the mostly minus convention is assumed here. Fermions are represented by Weyl spinors.
\bibitem{phi} We use a compact notation where $\phi$ is an array of real scalars, which includes all scalars of the theory.
\end{thebibliography}
\end{document}